\documentclass{article}
\usepackage{spconf}
\usepackage{algorithm,algpseudocode,amsfonts,amsmath,amssymb,amsthm,cite,enumitem,graphicx,hyperref,pgfplots,stmaryrd,subfig}
\usepackage{dsfont}
\usepackage{siunitx}
\usepackage{booktabs}
\usepackage{threeparttable,balance}
\newlist{todolist}{itemize}{2}
\setlist[todolist]{label=$\square$}
\hypersetup{    
colorlinks=true,
citecolor=blue(ryb),
linkcolor=crimson,
urlcolor=magenta
}
\PassOptionsToPackage{hyphens}{url}
\definecolor{azure}{rgb}{0.0, 0.5, 1.0}
\definecolor{frenchblue}{rgb}{0.0, 0.45, 0.73}
\definecolor{forestgreen(traditional)}{rgb}{0.0, 0.27, 0.13}
\definecolor{mygreen}{rgb}{0.09, 0.45, 0.27}
\definecolor{myblue}{rgb}{0.2383,0.5195,0.7734}
\definecolor{mygreen}{rgb}{0.6445,0.9297,0.0039}
\definecolor{darklavender}{rgb}{0.45, 0.31, 0.59}
\definecolor{americanrose}{rgb}{1.0, 0.01, 0.24}
\definecolor{pigblue}{rgb}{0.2, 0.2, 0.6}
\definecolor{blue(ryb)}{rgb}{0.01, 0.28, 1.0}
\definecolor{amethyst}{rgb}{0.6, 0.4, 0.8}
\definecolor{deepmagenta}{rgb}{0.8, 0.0, 0.8}
\definecolor{carminered}{rgb}{1.0, 0.0, 0.22}
\definecolor{iris}{rgb}{0.35, 0.31, 0.81}
\newcommand{\minimize}{\operatornamewithlimits{minimize}}
\newcommand{\st}{\operatornamewithlimits{subject \ to}}

\newcommand{\reals}{{\mbox{\bf R}}}

\newcommand{\prox}{\boldsymbol{prox}}

\def \diag    {\operatorname{diag} }

\def \reals    {{\mathbb R}}

\def\ccalF{{\ensuremath{\mathcal F}}}

\def\ccalO{{\ensuremath{\mathcal O}}}

\def\ccal0{{\ensuremath{\mathcal 0}}}
\def\bbA{{\ensuremath{\boldsymbol A}}}
\def\bbB{{\ensuremath{\boldsymbol B}}}
\def\bbC{{\ensuremath{\boldsymbol C}}}

\def\bbG{{\ensuremath{\boldsymbol G}}}
\def\bbH{{\ensuremath{\boldsymbol H}}}

\def\bbN{{\ensuremath{\boldsymbol N}}}

\def\bbU{{\ensuremath{\boldsymbol U}}}
\def\bbV{{\ensuremath{\boldsymbol V}}}
\def\bbS{{\ensuremath{\boldsymbol S}}}

\def\bbX{{\ensuremath{\boldsymbol X}}}
\def\bbY{{\ensuremath{\boldsymbol Y}}}
\def\bbZ{{\ensuremath{\boldsymbol Z}}}
\def\bbu{{\ensuremath{\boldsymbol u}}}
\def\bbv{{\ensuremath{\boldsymbol v}}}

\def\bbx{{\ensuremath{\boldsymbol x}}}

\def\bb0{{\ensuremath{\boldsymbol 0}}}
\def\hbH{{\hat{\ensuremath{\boldsymbol H}} }}

\def\hbY{{\hat{\ensuremath{\boldsymbol Y}} }}

\def\bbcalX{\mbox{\boldmath $\mathcal{X}$}}
\def\bbcalY{\mbox{\boldmath $\mathcal{Y}$}}

\definecolor{crimson}{rgb}{0.86, 0.08, 0.24}
\definecolor{scarlet}{rgb}{1.0, 0.13, 0.0}
\definecolor{hookersgreen}{rgb}{0.0, 0.44, 0.0} % good green
\definecolor{applegreen}{rgb}{0.55, 0.71, 0.0}
\definecolor{cultramarine}{rgb}{0.25, 0.0, 1.0}
\definecolor{ruddy}{rgb}{1.0, 0.0, 0.16}

\newcommand\blue[1]{{\color{frenchblue} #1}}

\def \fro {{\mathsf{F}}}
\def \tv {{\mathsf{TV}}}

\def \prox {\operatorname{prox}}

\def \inv {{-1}}

\title{Anisotropic Tensor Deconvolution of Hyperspectral Images}
\name{Xinjue Wang$^{1}$, Xiuheng Wang$^{2}$, Esa Ollila$^{1}$, Sergiy A. Vorobyov$^{1}$
\thanks{This research is supported by the Research Council of Finland under Grant 359848, and Grant 357715.}
}
\address{
$^{1}$Department of Information and Communications Engineering, Aalto University, Finland\\
$^{2}$ Université de Lorraine, CNRS, CRAN, France
}

\begin{document}% 
\maketitle
\vspace{-3mm}
% --- ABSTRACT ---
\begin{abstract}
Hyperspectral image (HSI) deconvolution is a challenging ill-posed inverse problem, made difficult by the data's high dimensionality.
We propose a parameter-parsimonious framework based on a low-rank Canonical Polyadic Decomposition (CPD) of the entire latent HSI $\bbcalX \in \reals^{P\times Q \times N}$.
This approach recasts the problem from recovering a large-scale image with $PQN$ variables to estimating the CPD factors with $(P+Q+N)R$ variables.
This model also enables a structure-aware, anisotropic Total Variation (TV) regularization applied only to the spatial factors, preserving the smooth spectral signatures.
An efficient algorithm based on the Proximal Alternating Linearized Minimization (PALM) framework is developed to solve the resulting non-convex optimization problem. 
Experiments confirm the model's efficiency, showing a numerous parameter reduction of over two orders of magnitude and a compelling trade-off between model compactness and reconstruction accuracy.

\end{abstract}
% -=-=-=-=-=-=-=-=-=-=-=-=-=-=-=-=-=-=-=-=-=-=-=-=-=-=-=-=-=-=-=-

%-=-=-=-=-=-=-=-=-=-=-=-=-=-=-=-=-=-=-=-=-=-=-=-=-=-=-=-=-=-=-=-
%-=-=-=-                Introduction                     -=-=-=-
%-=-=-=-=-=-=-=-=-=-=-=-=-=-=-=-=-=-=-=-=-=-=-=-=-=-=-=-=-=-=-=-
\section{Introduction}
% \cite{wang2020hyperspectral}
% {\sf
% \underline{To Do:}
% \begin{todolist}
%     \item Why tensor deconvolution is important in image/signal processing
%     \item Why CPD is used to model clean signal $\bbX$, and mention capturing low-rankness, sparsity and etc
%     \item Why TV, or mixed regularization, why not try PnP
%     \item Why factor C does not need TV regularization
%     \item How to choose $(\lambda)$
%     \item Convergence analysis
%     \item Limitation of current techniques used in tensor deconvolution, like high-computational complexity \red{(compare the computational complexity of WLRTR (a weighted tensor nuclear norm)~\cite{chang2020weighted} and 3DTFV (a 3D fractional total variation norm)~\cite{guo2021three} to support this point.)}
%     \item A deep unrolling architecture: Transformer,
% \end{todolist}
% }
 
High-dimensional data, such as hyperspectral images (HSIs) are often represented as tensors~\cite{sidiropoulos2017tensor,xu2013bcd_siam}. 
In practical acquisition systems, the signals are usually degraded by a convolution process (blurring) and corrupted by additive noise~\cite{henrot2012fast, wang2020learning, wang2023tuning}.
The task of \textit{tensor deconvolution} is to recover the latent clean signal $\bbcalX\in \reals^{P\times Q \times N}$ from its degraded observation $\bbcalY$ is a fundamental ill-posed inverse problem.
% This problem, often referred to as \textit{hyperspectral image deconvolution}, presents the challenge of accounting for the reduction in the number of estimated parameters for model compactness.
% videos, and Magnetic resonance imaging data, 
% This is a fundamental inverse problem in signal and image processing.
% Recently, it has attracted attention to restore HSI data degraded by blurring and noise~\cite{henrot2012fast, wang2020learning, wang2023tuning}. 

%  ---
% {\it Related work.}
Different prior information on the latent images has been considered to regularize the solution in the HSI deconvolution task. For example, the fast hyperspectral restoration algorithm in~\cite{henrot2012fast} performs deconvolution under positivity constraints
while accounting for spatial and spectral correlation. In~\cite{song2019online}, an online deconvolution algorithm is devised by considering sequentially
collected data by push-broom devices. 3D fractional total variation (3DFTV)~\cite{guo2021three} leverages non-local smoothness in all dimensions to better preserve textures. 
More recently, learning-based regularization methods, such as plug-and-play~\cite{wang2020learning, wang2023tuning} and deep unrolling~\cite{gkillas2023highly} frameworks, have shown superior performance in reconstruction accuracy. 
% For example, the hyper-Laplacian prior (HLP)~\cite{krishnan2009fast_LaplacianPrior} promotes gradient sparsity, 
% spatial and spectral priors (SSP)~\cite{henrot2012fast} enforces smoothness in both spatial and spectral domains and 
% 3-D fractional total variation (3DFTV)~\cite{guo2021three} leverages non-local smoothness to better preserve textures.  
%

However, these are full-rank models that operate on the entire tensor, and thus must solve a high-dimensional optimization problem without addressing the data's intrinsic low-rank structure.
A different line of work leverages the low-rankness of the latent images.
For example, the weighted low-rank tensor recovery method (WLRTR)~\cite{chang2020weighted} leverages the non-local self-similarity within the HSI by searching for similar 3D patches and assuming that the resulting patch-group forms a low-rank tensor. 
However, this patch-based strategy retains the full-rank tensor as its optimization variable and relies on computationally expensive patch processing that neglects the scene's global structure.
% The limitations of the existing approaches motivate our direct, global modeling paradigm.

% ---
{\it Contributions.}
In this work, we propose a novel tensor deconvolution framework for HSIs. 
Our framework is composed of three key components: \textit{(i)} a low-rank CPD model that recasts the recovery problem from a large-scale image recovery problem to a small-scale parameter estimation problem; \textit{(ii)} a structure-aware, anisotropic Total Variation regularization applied to the spatial factors, preserving the smooth signatures; and \textit{(iii)} an efficient algorithm based on the PALM framework~\cite{bolte2014PALM, CosseratPesquet2024IVA_convergence} to solve the resulting non-convex problem.
The source code is available at \url{https://github.com/xnnjw/TensorDeconv_release}.

{\it Notation.}
Boldface lower case letters such as $\bbx$ represent column vectors, boldface capital letters like $\bbX$ denote matrices, and boldface capital calligraphic letters like $\bbcalX$ denote tensors.
% The 2D convolution is denoted by $\star$, and the Hadamard product is expressed with the symbol $\odot$.
% Let $[\cdot]_+$ be the projection operator onto the non-negative orthant. 
Discrete Fourier Transform (DFT) is denoted by $\ccalF\{\cdot\}$, and $\ccalF^{-1}\{\cdot\}$ denotes the inverse DFT (IDFT).
We use $\diag(\cdot)$ to denote a diagonal matrix constructed by a vector.
The notation $\bbX(i,\cdot)$ is used to denote the $i$-th row of $\bbX$, and $\bbX(\cdot,j)$ is used to denote the $j$-th column of $\bbX$.

%-=-=-=-=-=-=-=-=-=-=-=-=-=-=-=-=-=-=-=-=-=-=-=-=-=-=-=-=-=-=-=-
%-=-=-=-               Problem Formulation               -=-=-=-
%-=-=-=-=-=-=-=-=-=-=-=-=-=-=-=-=-=-=-=-=-=-=-=-=-=-=-=-=-=-=-=-
\section{Problem Formulation}
\label{sec:PF}
% ------------
\subsection{Tensor Imaging Model}

We consider the problem of constructing a clear HSI $\bbcalX \in \reals^{P\times Q \times N}$ from its degraded observation $\bbcalY\in\reals^{P\times Q \times N}$, where $P, Q$, and $N$ are the number of rows, columns, and spectral bands.
The ill-posed nature of the inverse problem necessitates the use of effective priors. A powerful prior is the low-rankness of the latent images, which has been widely used in the literature~\cite{veganzones2015hyperspectral, chang2020weighted, borsoi2024personalized}. 
Based on this prior information, we consider modeling $\bbcalX$ directly as a single, global, low-rank tensor via the Canonical Polyadic Decomposition (CPD)~\cite{carroll1970analysis}, also known as the Kruskal model~\cite{Kruskal1977kruskal}. 
The CPD factorizes the entire tensor into a sum of $R$ rank-one components, where $R$ is the rank for CPD.
This factorization is defined by factor matrices $\bbA \in \reals^{P \times R}$, $\bbB \in \reals^{Q \times R}$, and $\bbC \in \reals^{N \times R}$ as
\begin{align}
    \label{eq:cpddef}
    \bbcalX \approx \llbracket \bbA, \bbB, \bbC \rrbracket = \sum_{r=1}^R \bbA(\cdot,r) \circ \bbB(\cdot,r) \circ \bbC(\cdot,r),
\end{align}
where $\circ$ denotes the vector outer product, $\bbA$ and $\bbB$ correspond to spatial abundances, and $\bbC$ corresponds to spectral signatures.
% These factors correspond to spatial abundances $\bbA, \bbB$ and spectral signatures~$\bbC$.
% This representation is physically meaningful for HSI, where the
Furthermore, it is parameter-parsimonious, reducing the number of variables to estimate from $PQN$ for the full tensor to $(P+Q+N)R$ for the factor matrices.
The deconvolution is thus reformulated from a high-dimensional image recovery problem into a lower-dimensional parameter estimation problem.

% ---
% Let $\bbcalX \in \reals^{ P \times Q \times N}$ and $\bbcalY \in \reals^{ P \times Q \times N}$ be the latent clean hyperspectral image and its degraded observation.
The degradation process is modeled on a slice-by-slice basis, i.e., the 3D tensor $\bbcalX$ is processed by its frontal slices $\{\bbX_i\}_{i=1}^N$ with $\bbX_i\in \reals^{P \times Q}$.
Each observed slice $\bbY_i \in \reals^{P \times Q}$ is assumed to be a 2D convolution of the corresponding clean slice $\bbX_i$ with a blurring kernel $\bbH_i\in \reals^{P \times Q}$, corrupted by additive independent and identically distributed (i.i.d.) Gaussian noise $\bbN_i\in \reals^{P \times Q}$ as
\begin{align}
    \label{eq:degradation_model}
    \bbY_i = \bbH_i \star \bbX_i + \bbN_i, \quad i=1,\ldots,N,
\end{align}
where $\star$ denotes the 2D convolution.
%
% We model the latent tensor $\bbcalX$ using a low-rank CPD of rank $R$~\eqref{eq:cpddef}.
The clean slice $\bbX_i$ in the degradation model \eqref{eq:degradation_model} is thus constructed from the CPD factor matrices $\bbA $, $\bbB$, and $\bbC$ as
\begin{align}
\label{eq:slice_construction}
\bbX_i = \sum_{r=1}^R \bbC(i,r) \bbA(\cdot,r)\bbB(\cdot,r)^\top.
\end{align}
The goal of the deconvolution is to recover the factor matrices $\bbA, \bbB, \bbC$ from the observations $\bbcalY$.

% --------
\subsection{The optimization problem}
\label{subsec:propsoedmethod:optprob}
Based on the degradation model~\eqref{eq:degradation_model} and the CPD representation~\eqref{eq:slice_construction}, the restoration of the latent tensor $\bbcalX$ is equivalent to estimating its factor matrices $\bbA, \bbB, \bbC$ from the observations $\bbcalY$. Thus, we can formulate the tensor deconvolution of HSIs as the following constrained optimization problem:
\begin{equation}
\label{eq:tensor_deconv1}
\begin{aligned}
    \minimize_{\bbA, \bbB, \bbC} \quad &
    % F(\bbA, \bbB, \bbC) := 
    f(\bbA, \bbB, \bbC) + 
    g_A(\bbA) + g_B(\bbB), \\
    \st \quad & \bbA\geq 0, \bbB \geq 0, \bbC \geq 0.
\end{aligned}
\end{equation}
The objective function consists of a smooth data fidelity term, $f(\bbA, \bbB, \bbC)$, and non-smooth regularization terms $g_A(\bbA)$, $g_B(\bbB)$ which encode prior information of the latent tensor~$\bbcalX$.
The non-negative constraints are imposed to reflect physical constraints, such as material abundances and spectral signatures being non-negative.
% The non-negativity constraints on the factor matrices $\bbA, \bbB, \bbC$ are imposed to reflect their physical meaning, such as material abundances and spectral signatures being non-negative.
% -
The data fidelity term $f(\bbA, \bbB, \bbC)$ is defined as
% ---
\begin{align}
    f(\bbA, \bbB, \bbC) = & \frac12 \sum_{i=1}^N \|\bbY_i - \bbH_i \star (\bbA \diag(\bbC(i,:)) \bbB^\top) \|_{\fro}^2 \notag \\
    & + \lambda_1 \|\bbA\|_{\fro}^2 + \lambda_2 \|\bbB\|_{\fro}^2 + \lambda_3 \|\bbC\|_{\fro}^2, \label{eq:smooth_f}
\end{align}
which combines the least-square (LS) loss with zero-order Tikhonov (Frobenius-norm) penalties on the factors to improve conditioning and numerical stability.
Here, $\lambda_1,\lambda_2,\lambda_3$ are non-negative hyperparameters.
The non-smooth terms regularize the spatial factors using the TV norm~\cite{rudin1992TV}, defined as
\begin{align}
    g_A(\bbA) = \lambda_A \sum_{r=1}^R \|\bbA_r\|_\tv,
    \ g_B(\bbB) = \lambda_B \sum_{r=1}^R \|\bbB_r\|_\tv, \label{eq:gAB}
\end{align}
where $\lambda_A \geq 0$ and $\lambda_B \geq 0$ control the strength of the regularization, with the TV norm for a vector $\bbv \in \reals^{P}$ defined~as
\begin{align}
    \|\bbv\|_\tv = \sum_{p=1}^{P-1} | \bbv_{p+1} - \bbv_{p} |.
\end{align}
This encourages piecewise constant patterns, preserving sharp transitions between regions~\cite{rudin1992TV}.

\subsection{PALM Framework}
The PALM algorithm~\cite{bolte2014PALM} is an effective framework for problems of the form:
\begin{align*}
    \min_{x_1, \dots, x_p} F(x_1, \dots, x_p) = \sum_{i=1}^p g_i(x_i) + f(x_1, \dots, x_p),
\end{align*}
where each $g_i(x_i)$ is a (possibly non-convex) non-smooth function, and $f(x_1, \dots, x_p)$ is a smooth coupling function.

The core of PALM is a block coordinate descent scheme that sequentially minimizes a surrogate function for each variable block.
At iteration $k$, to update block $x_i$, PALM linearizes the smooth term $f$ at the current iterate and adds a quadratic proximal term. 
This leads to the update rule
\begin{align*}
    x_i^{k+1} \in \arg\min_{u} \ g_i(u) + \langle \nabla_i f, u - x_i^k \rangle + \frac{1}{2c_{i,k}} \|u - x_i^k\|_2^2 ,
\end{align*}
where $\nabla_i f$ is the partial gradient of $f$ with respect to $x_i$ at~$x_i^k$, and $c_{i,k} > 0$ is a step size. 
This update can be written more compactly as a proximal gradient step. 
The step size $c_{i,k}$ is crucial for convergence and can be found using a backtracking line search method.

\section{Proposed method}
\label{sec:Proposed_method}
% ---
% \textit{Proximal Operator for Total Variation.}
Solving the optimization problem \eqref{eq:tensor_deconv1} involves the proximal operators of $g_A$ and $g_B$ \eqref{eq:gAB}.
A key property of these operators is their separability: since the regularization is a sum of TV norms applied independently to each column, the overall proximal operator can be computed by applying the one-dimensional (1D) TV proximal operator to each column of the input matrix.
For an input matrix $\bbU = [\bbu_1,\ldots,\bbu_R]$, this separability is expressed~as
\begin{equation}
    \begin{split}
        & \minimize_{\bbv_1,\ldots,\bbv_R} \quad \sum_{r=1}^R \left( \|\bbv_r\|_\tv + \frac12 \|\bbv_r - \bbu_r\|_2^2 \right) \\
    & \equiv
    \minimize_{\bbv_1,\ldots,\bbv_R} \quad \sum_{r=1}^R \|\bbv_r\|_\tv + \frac12 \|\bbV - \bbU\|_\fro^2.
    \end{split}
\end{equation}
This separability is computationally advantageous. The core task of computing the 1D proximal operator $\prox_{\|\cdot\|_\tv}(\cdot)$ can be efficiently solved~\cite{Chandler2025AugClosedTV}.

Our choice of regularizers is tailored to the distinct physical meaning of each factor matrix in the CPD model.
For the factor matrices $\bbA$ and $\bbB$, which represent piecewise-constant spatial abundance maps, TV is a proper regularization for preserving their sharp edges.
Conversely, the columns of matrix $\bbC$ represent smooth spectral signatures, for which TV regularization may introduce artifacts. 
This tailored TV regularization, combined with Frobenius norm penalties for numerical stability and non-negativity constraints to preserve physical meaning, allows our model to better reflect the intrinsic properties of HSI.

%-=-=-=-=-=-=-=-=-=-=-=-=-=-=-=-=-=-=-=-=-=-=-=-=-=-=-=-=-=-=-=-
%-=-=-=-               Proposed Algorithm                -=-=-=-
%-=-=-=-=-=-=-=-=-=-=-=-=-=-=-=-=-=-=-=-=-=-=-=-=-=-=-=-=-=-=-=-
\subsection{PALM for tensor deconvolution}

% ------------------- Algorithm BEGIN -------------------
\begin{algorithm}[!t]
    \caption{Proposed PALM for solving~\eqref{eq:tensor_deconv1}}
    \begin{algorithmic}[1]
    
    \State\textbf{Input:} $\hbY_i = \ccalF\{\bbY_i\}, \hbH_i = \ccalF\{\bbH_i\}, \forall i$.

    \State\textbf{Initialize:} 
    $\bbA^{0}$, $\bbB^{0}$, $\bbC^{0}$, initial step sizes $c_0, d_0, e_0 = 1$, backtracking parameter $\beta, \eta \in (0,1)$, and stopping threshold $\epsilon=10^{-6}$.
    % \State\textbf{Pre-compute:} 
    % $\forall i, \ \hbY_i = \ccalF\{ \bbY_i \},\hbH_i = \ccalF\{ \bbH_i \}$
    
    % \For{$k=1,2,\ldots$}
    \For{$k=0,1,2,\ldots$}
    % 1/3 A
    \Statex \blue{// Update block $\bbA$} 
    \State $\bbG_A \leftarrow \nabla_A f(\bbA^k, \bbB^k, \bbC^k)$
    \State $c \leftarrow c_k / \eta$ \Comment{Initial trial step size for this iteration}
    \State $(\bbA^{k+1}, c_{k+1}) \gets \textsf{BacktrackLS}(\bbA^k, \bbG_A, c, \beta)$ 
    % \Comment{Fixed blocks $\bbB^k, \bbC^k$}
    
    % 2/3 B
    \Statex \blue{// Update block $\bbB$} 
    \State $\bbG_B \leftarrow \nabla_B f(\bbA^{k+1}, \bbB^k, \bbC^k)$
    \State $d \leftarrow d_k / \eta$
    \State $(\bbB^{k+1}, d_{k+1}) \gets \textsf{BacktrackLS}(\bbB^k, \bbG_B, d, \beta)$ 
    % \Comment{Fixed blocks $\bbA^{k+1}, \bbC^k$}
    
    % 3/3 C
    \Statex \blue{// Update block $\bbC$}
    \State $\bbG_C \leftarrow \nabla_C f(\bbA^{k+1}, \bbB^{k+1}, \bbC^k)$
    \State $e \leftarrow e_k / \eta$
    \State $(\bbC^{k+1}, e_{k+1}) \gets \textsf{BacktrackLS}(\bbC^k, \bbG_C, e, \beta)$ 
    % \Comment{Fixed blocks $\bbA^{k+1}, \bbC^{k+1}$}
    %
    \If{Stopping criteria is met}
        \State \textbf{return} $\bbA^{k+1}$, $\bbB^{k+1}$, $\bbC^{k+1}$
    \EndIf
    \EndFor
    \end{algorithmic}
    \label{alg:PALM_tensordeconv}
\end{algorithm}
% ------------------- Algorithm END -------------------

% ------------------- Algorithm BEGIN -------------------
\begin{algorithm}[!t]
\caption{{\sf BacktrackLS}: Backtracking Line Search}
\label{alg:backtracking}
\begin{algorithmic}[1]
\State \textbf{Input:} Current iterate $\bbZ^k$, gradient $\bbG_Z$, regularizer $g_Z$, initial step size $t$, factor $\beta \in (0,1)$.
\Repeat
    \State $\bbU \gets \big[\prox_{t g_Z}(\bbZ^k - t \bbG_Z)\big]_+$ \label{alg2:update_step}
    \State $f_{\rm LHS} \gets f(\dots, \bbU, \dots)$
    \State $f_{\rm RHS} \gets f(\dots, \bbZ^k, \dots) + \langle \bbG_Z, \bbU - \bbZ^k \rangle + \frac{1}{2t} \|\bbU - \bbZ^k\|_\fro^2$
    \State $t \gets \beta t$ \Comment{Shrink step size}
    % \If{$f_{\rm LHS} \le f_{\rm RHS}$}
    %     \State \textbf{return} $t, \bbU$
    % \Else
    %     \State $t \gets \beta t$ \Comment{Shrink step size}
    % \EndIf
\Until{$f_{\rm LHS} \le f_{\rm RHS}$}
\State $t_{k+1}\gets t/\beta$
\State $\bbZ^{k+1} \gets \bbU$
\State {\bf Return:} $t_{k+1}, \bbZ^{k+1}$
\end{algorithmic}
\end{algorithm}
% ------------------- Algorithm END -------------------

To solve the constrained optimization problem~\eqref{eq:tensor_deconv1}, we propose an algorithm based on the PALM framework. 
The method cyclically updates each factor matrix by minimizing a linearized surrogate function of the objective, subject to the non-negativity constraints.
This procedure results in a projected proximal gradient step for each block.
Let $[\cdot]_+$ denote the projection operator onto the non-negative orthant.
The updates for the three factors at iteration $k$ are given as follows
\begin{enumerate}[itemsep=0pt, topsep=0pt, parsep=0pt, partopsep=0pt]
    \item $\bbA^{k+1} = \big[ \prox_{c_k g_A} \big( \bbA^k - c_k \nabla_\bbA f(\bbA^k, \bbB^{k}, \bbC^{k}) \big) \big]_+$
    \item $\bbB^{k+1} = \big[ \prox_{d_k g_B} \big( \bbB^k - d_k \nabla_\bbB f(\bbA^{k+1}, \bbB^k, \bbC^{k}) \big) \big]_+$
    \item $\bbC^{k+1} = \big[ \bbC^k - e_k \nabla_\bbC f(\bbA^{k+1}, \bbB^{k+1}, \bbC^k) \big]_+$
\end{enumerate}
The proximal operators for $g_\bbA(\bbA)$ and $g_\bbB(\bbB)$ handle the TV regularization as detailed in Section~\ref{subsec:propsoedmethod:optprob}, while the update for $\bbC$ simplifies to a standard projected gradient step. 
The full procedure is detailed in Algorithm~\ref{alg:PALM_tensordeconv}.\
It proceeds by iteratively updating the three factors, where each update involves a backtracking line search to determine the step size.

% ---
{\it Backtracking line search.}
The goal of the line search is to find a step size (e.g., $c_k$ for block $\bbA$) that satisfies the sufficient decrease condition derived from the PALM framework
\begin{align}
    \label{eq:sufficient_decrease_condition_A}
    & f(\bbA^{k+1}, \bbB^k, \bbC^k) \le f(\bbA^k, \bbB^k, \bbC^k) \\
    & + \langle \nabla_\bbA f(\bbA^k, \bbB^k, \bbC^k), \bbA^{k+1} - \bbA^k \rangle + \frac{1}{2c_k} \|\bbA^{k+1} - \bbA^k\|_\fro^2, \notag
\end{align}
This procedure is encapsulated in our generic backtracking module, {\sf BacktrackLS}, detailed in Algorithm~\ref{alg:backtracking}.

The generic update in line~\ref{alg2:update_step} of Algorithm~\ref{alg:backtracking} adapts to each block. For blocks $\bbA$ and $\bbB$, $g_Z$ corresponds to the TV regularizers~\eqref{eq:gAB}.
For block $\bbC$, the proximal operator $\prox_{t g_C}(\cdot)$ becomes the identity operator, and the update simplifies to $\bbU \gets [\bbC^k - t\bbG_C]_+$.
The function evaluations $f$ implicitly use the most recently updated values for the other fixed blocks.

% % -=-=-=--=-=-=--=-=-=--=-=-=--=-=-=-
% \vspace{3mm}
{\it Gradient Computation.}
The update steps in Algorithm~\ref{alg:PALM_tensordeconv} require the partial gradients of the smooth term $f$. 
The main computational bottleneck lies in the 2D convolutions within $f$, which we accelerate significantly by performing operations via DFT.
We first calculate an intermediate term $\bbS_i$ for each slice $i$ at every iteration $k$:
\begin{equation}
\label{eq:Iterative_S_compact}
\bbS_i^{k} = \ccalF^\inv \{ \hbH_i \odot (\hbH_i \odot \ccalF \{\bbX_i^{k}\} - \hbY_i) \}.
\end{equation}
where $\odot$ denotes the Hadamard product, $\bbX_i^k$ is the current estimate of the $i$-th clean slice, and $\hbY_i$ and $\hbH_i$ are the pre-computed DFTs of the observation and blur kernel, respectively.
The partial gradients of $f$~\eqref{eq:smooth_f}~are
% \begin{enumerate}[itemsep=0pt, topsep=0pt, parsep=0pt, partopsep=0pt]
%     \item $\nabla_{\bbA} f = \sum_{i=1}^{N} \bbS_i \bbB \diag(\bbC(i,:)) + 2\lambda_1 \bbA$
%     \item $\nabla_{\bbB} f = \sum_{i=1}^{N} \bbS_i^\top \bbA \diag(\bbC(i,:)) + 2\lambda_2 \bbB$
%     \item $\nabla_{\bbC} f(i,:) = \diag(\bbA^\top \bbS_i \bbB)^\top + 2\lambda_3 \bbC(i,:)$
% \end{enumerate}
%
\begin{subequations}
\begin{align}
\nabla_{\bbA} f & = \sum_{i=1}^{N} \bbS_i \bbB \diag(\bbC(i,\cdot)) + 2\lambda_1 \bbA, \label{eq:grad_f_A} \\
\nabla_{\bbB} f & = \sum_{i=1}^{N} \bbS_i^\top \bbA \diag(\bbC(i,\cdot)) + 2\lambda_2 \bbB,
\label{eq:grad_f_B}\\
\nabla_{\bbC} f(i,\cdot) & = \diag(\bbA^\top \bbS_i \bbB)^\top + 2\lambda_3 \bbC(i,\cdot).
\label{eq:grad_f_C}
\end{align}
\end{subequations}
Leveraging the DFT, the overall computational complexity per iteration is dominated by the slice reconstructions~\eqref{eq:slice_construction} and DFTs, scaling as $\ccalO(N(PQR+PQ\log(PQ)))$, which in practice simplifies to $\ccalO(NPQR)$ as the rank $R$ is typically larger than $\log(PQ)$.

%-=-=-=-=-=-=-=-=-=-=-=-=-=-=-=-=-=-=-=-=-=-=-=-=-=-=-=-=-=-=-=-
%-=-=-=-                   Experiments                   -=-=-=-
%-=-=-=-=-=-=-=-=-=-=-=-=-=-=-=-=-=-=-=-=-=-=-=-=-=-=-=-=-=-=-=-
\section{Numerical Experiments}
\label{sec:Experiment}
% 
% \subsection{Set Up}

% ---
\begin{table}[!t]
\centering
% \caption{Quantitative comparison of the proposed and comparing methods on the CAVE dataset, in terms of the numbers of Parameters, Memory and reconstruction error (RMSE, PSNR).}
\caption{Quantitative comparison against reference methods on the CAVE dataset in terms of model complexity and reconstruction error.}
\resizebox{\linewidth}{!}{
\begin{tabular}{lccccc}
% \toprule
\hline
\hline
\textbf{Methods} & \# \textbf{Paras.} & \textbf{Mem.} (MB) & \textbf{RMSE} & \textbf{PSNR} \\
\midrule
HLP     & $\sim 8 \times 10^6$ & 62 & 4.42 & 36.17 \\
SSP     & $\sim 8 \times 10^6$ & 62 & 4.85 & 35.37 \\
WLRTR   & $\sim 8 \times 10^6$ & 62 & 4.74 & 35.87 \\
3DFTV   & $\sim 8 \times 10^6$ & 62 & 4.33 & 36.45 \\
\hline
Proposed    & $\sim 3 \times 10^4$ & 0.24 & 6.99 & 32.94 \\
% \bottomrule
\hline
\hline
\end{tabular}}
\label{tab:comparison}
\vspace{-3mm}
\end{table}

% ---
\begin{figure}[!t]
    \centerline{\includegraphics[width = 0.7\linewidth]{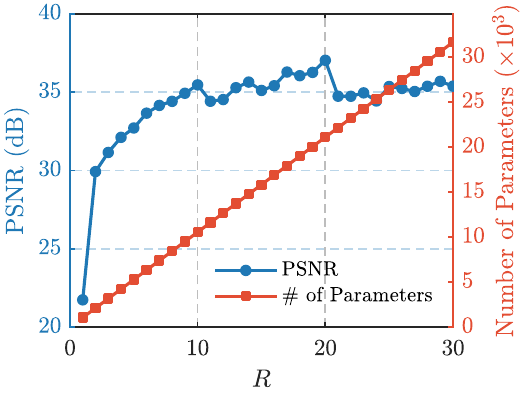}}
    \vspace{-3mm}
    \caption{
    Best PSNR value (left axis) and number of parameters (right axis) versus CPD rank $R$ on an image from the CAVE dataset. } 
    \label{fig:PSNRvsRank}
    \vspace{-5mm}
\end{figure}
% ---

Now we turn to evaluate our proposed method against several baselines: hyper-Laplacian prior (HLP)~\cite{krishnan2009fast_LaplacianPrior}, spatial-spectral prior (SSP)~\cite{henrot2012fast}, WLRTR~\cite{chang2020weighted}, and 3DFTV~\cite{guo2021three}.
The experiments are conducted on the Columbia Multispectral Database (CAVE)\footnote{\url{https://cave.cs.columbia.edu/repository/Multispectral}}~\cite{yasuma2010generalized}. 
This dataset contains 32 HSIs of size $512\times512\times31$ $(P=512, Q=512, N=31)$, which we scale to the interval [0,1] to serve as the ground-truth clean images $\bbcalX$.
The degraded observations $\bbcalY$ are then simulated from the ground truth. 
First, the clean image is blurred using a spectrally invariant 2D Gaussian kernel in each spectral band. 
The kernel size is set to $9\times 9$ pixels with a standard deviation of $2$.
Second, additive white Gaussian noise with a standard deviation of $0.01$ is added to each band.
% For the convolution operation, this small kernel is zero-padded to match the full image dimensions. 

Table~\ref{tab:comparison} presents the quantitative results. 
The proposed low-rank model reduces the number of parameters by over two orders of magnitude (from around $8$ million to $30,000$) and the memory by a similar factor. 
This highlights the core advantage of the proposed model, which recasts the large-scale image recovery task into a much more tractable parameter estimation problem.
In contrast, the competing full-rank methods achieve lower root-mean-square error (RMSE) and higher peak-signal-to-noise ratio (PSNR) values. 
Our method thus offers a trade-off between model parsimony and reconstruction accuracy, making it an alternative for resource-constrained applications.

Fig.~\ref{fig:PSNRvsRank} plots the number of model parameters and the best PSNR value against the CPD rank $R$ for the ``real and fake peppers" image of the CAVE dataset.
The number of parameters scales as $(P+Q+N)R$ and grows linearly with the rank $R$.
The PSNR value initially increases with $R$, but peaks at an optimal rank ($R=20$ in this example) before plateauing.
The existence of such an optimal rank suggests that the underlying data possesses a low-rank structure; otherwise, the PSNR would likely increase monotonically as a higher-rank model begins to overfit the noise. 
We note that determining the optimal tensor rank a priori is a well-known NP-hard problem~\cite{Hillar2013NPHardTensorrank} and is beyond the scope of this work.

%-=-=-=-=-=-=-=-=-=-=-=-=-=-=-=-=-=-=-=-=-=-=-=-=-=-=-=-=-=-=-=-
%-=-=-=-                   Conclusion                    -=-=-=-
%-=-=-=-=-=-=-=-=-=-=-=-=-=-=-=-=-=-=-=-=-=-=-=-=-=-=-=-=-=-=-=-
\section{Conclusion}
\label{sec:Conclusion}
In this paper, a novel framework for HSI deconvolution was proposed based on a low-rank CPD model. 
By leveraging the physical interpretation of the CPD factors, we designed an anisotropic Total Variation regularization scheme that targets only the spatial modes of the HSI. 
Experimental results demonstrated that our method is extremely parameter-efficient, reducing model complexity by over two orders of magnitude while maintaining competitive reconstruction performance. 

% \vfill
% \pagebreak

% \clearpage
% \balance
% \ninept
\newpage
%-=-=-=-=-=-=-=-=-=-=-=-=-=-=-=-=-=-=-=-=-=-=-=-=-=-=-=-=-=-=-=-
%-=-=-=-                   Reference                     -=-=-=-
%-=-=-=-=-=-=-=-=-=-=-=-=-=-=-=-=-=-=-=-=-=-=-=-=-=-=-=-=-=-=-=-
\bibliographystyle{IEEEtran}
\bibliography{IEEEabrv, reference, ref_HSI_deconv}
\end{document}